# The hidden sustainability bottleneck in high-entropy alloy design

**Alexandre Nominé**[1,2,3,4*], **Ayyoub El-Kasmi**[1,5], **Danielle Beaulieu**[1,3], **Outhmane Ouahri**[1,5], **Nguyen Thuy Huong**[1,6], **Wassim Amzil**[1,5], **Aymane Droussi**[1,5], **Oleksandra Kuksa**[1,2,7], **Eirini Katsarou**[1,8], **Chahrazed Labba**[1,2], **Anne Boyer**[1,2], **Hani Henein**[9], **Thierry Belmonte**[3], **François Rousseau**[1,3], **Tuncay Gurbuz**[10], **Elena Mitrofanova**[1,11], **Agnès Samper**[12], **Pascal Bolz**[12], **Valentin Milichko**[1,3,13], **Olga Chernoburova**[14], **Alexandre Chagnes**[14], **Michel Cathelineau**[14], **Uroš Cvelbar**[4,15*], **Janez Zavašnik**[4,16 *]

[1] *University of Lorraine, Mines Nancy, Nancy, France*
[2] *LORIA, University of Lorraine – INRIA - CNRS, Vandoeuvre lès Nancy, France*
[3] *Institut Jean Lamour, Université de Lorraine – CNRS, Nancy, France*
[4] *Department of Gaseous Electronics, Jožef Stefan Institute, Ljubljana, Slovenia*
[5] *EMINES, University Mohamed VI, Ben Guerrir, Morocco*
[6] *Hanoi University of Science and Technology, Hanoi, Viet-Nam*
[7] *Kyiv Polytechnic Institute, Kyiv, Ukraine*
[8] *Department of Physics, Aristotle University Thessaloniki, Thessaloniki, Greece*
[9] *Department of Chemical and Materials Engineering, University of Alberta, Edmonton, Canada*
[10] *Galatasaray Üniversitesi, Beşiktaş/İstanbul, Turkey*
[11] *Faculty of Physics, ITMO University, St Petersburg, Russia*
[12] *Sustainable Minerals Institute, The University of Queensland, Brisbane, Australia*
[13] *New Uzbekistan University, Tashkent, 100000, Uzbekistan*
[14] *University of Lorraine, CNRS, CREGU, GeoRessources, Nancy, France*
[15] *Jožef Stefan International Postgraduate School, Ljubljana, Slovenia*
[16] *Max-Planck-Institut für Nachhaltige Materialien, Düsseldorf, Germany*
[*] *email: janez.zavasnik@ijs.si ; uros.cvelbar@ijs.si, alexandre.nomine@univ-lorraine.fr*

**Abstract:**

Because of the enormous number of potential compositions comparable to the number of stars in the universe, high entropy alloys (HEAs) are a virtually endless source of materials possessing versatile properties. Among them, HEAs are promising substitutes for critical elements such as rare earths or platinum group metals. Random or incremental development methods are neither practical nor efficient for exploration. Targeted selection with sustainability in mind is a necessary enabler, but choosing the suitable sustainable composition of HEAs is challenging. In this study, we perform a comprehensive sustainability assessment of 30201 HEAs and identify a resilient shortlist (≈5%) of compositions with consistently high sustainability profiles, providing a prioritized set for future experimental investigation. We consider various sustainability criteria such as carbon footprint,

**ESG ratings, production compatibility levels and reserves. The results offer a strategic roadmap for HEA scientists, guiding their experimental efforts towards the most sustainable compositions, supporting industrial sustainability goals while optimising the use of time and resources.**

The green and digital transitions are intrinsically tied to metals, which are non-renewable resources[1]. The extraction and use of these metals have multifaceted impacts, encompassing environmental, social, and geopolitical dimensions[2]. As these transitions progress, there is growing concern regarding potential bottlenecks in the global metal supply chains[3].

To effectively address this challenge, a dual strategy is necessary: enhancing recycling processes, and developing alternative materials to replace the most critical ones. While recycling is undeniably crucial, it alone will not suffice to meet the increasing demand for materials driven by digital and green transitions[4,5]. Moreover, while demand for raw materials is rising, the exploration budget for critical metals is at its lowest point[6]. This paradox poses a significant risk to the supply of crucial metals for the twin transition in the upcoming decade[7]. To mitigate overconsumption and accelerate the transformation of society towards a circular economy, more attention should be placed on innovative product design that balances cost-effectiveness, performance and sustainability[8]. Achieving this objective demands urgent and extensive efforts to discover new materials with advanced functional properties and high sustainability standards.

High Entropy alloys (HEAs) are characterized by their unique compositional design consisting of multiple principal elements in near-equimolar proportions, presenting opportunities to overcome traditional trade-offs in both mechanical[9–11] or chemical[12,13] properties. These advancements are particularly noteworthy, considering that only a tiny fraction of the vast compositional space has been explored[14]. The application of Artificial Intelligence has significant potential to accelerate the discovery of new HEAs[15]. However, such an approach remains intrinsically dependent on experimental data, which are expensive and time-consuming to generate. Therefore, to expedite the development of substitute materials, our experimental discovery efforts must be guided by a dual focus: combining high performance with long-term environmental and economic sustainability.

Researchers and governments have developed diverse approaches to metals criticality assessment[16,17], incorporating geological, social, geopolitical, and environmental considerations. Recent works by Graedel et al. and Fu et al.[18–20] provide foundations for criticality and sustainability assessment, which we build upon here by extending these frameworks into large-scale HEA design. Building on these frameworks, sustainable materials design principles have recently been extended to High Entropy Alloys (HEAs), though previous studies have typically examined datasets of several hundred to a few thousand experimentally synthesized alloys[21,22]. In this study, we analyse the 30201 alloys identified by Chen *et al.*[23] as potential HEAs from among 658000 possible equimolar compositions derived from a palette of 40 elements. These candidate alloys are assessed using a comprehensive set of sustainability criteria, including $CO_2$ emissions, energy consumption, environmental impact, social factors, and governmental risks[2], enhanced by incorporating the often-overlooked factor of *companionality*[24]. Based on these factors, we develop a sustainability risk score, which, combined with production metrics and reserve scarcity, can serve as a guiding framework for evaluating the substitution potential of HEA systems, which is highly dependent on the intended market size. Our evaluation yields a ranked list of the 1556 HEAs showing the most favourable sustainability profiles.

**Embodied energy and $CO_2$ footprint of metal extraction**

Metal extraction and processing are highly energy-intensive processes that significantly contribute to global $CO_2$ emissions. As shown in **Figure 1a**, the $CO_2$ footprint of metal extraction varies dramatically across different elements (See **Section 1** of the *Supplementary Information File*). Rock-forming elements (Fe, Al, Si, Mn, Mg, Ti) have relatively low carbon footprints, typically below 10 kg $CO_2$/kg. Refractory elements (Cr, Nb, Mo, Hf, Ta, W) have intermediate values, while Platinum Group Metals (PGMs; Pt, Pd, Ir, Os, Rh, Ru) demonstrate very high carbon intensities, reaching up to several tons of $CO_2$/kg for elements like Osmium.

The figure also reveals substantial disparities in recycling rates across different metal groups. While some elements have recycling rates above 30%, many critical metals have negligible recycling rates. The recycling footprint, also shown in **Figure 1a**, demonstrates substantial potential for $CO_2$ reductions compared to primary production. The actual carbon footprint is calculated as a weighted average of the primary and secondary (recycled) footprints, based on each element's specific recycling rate.

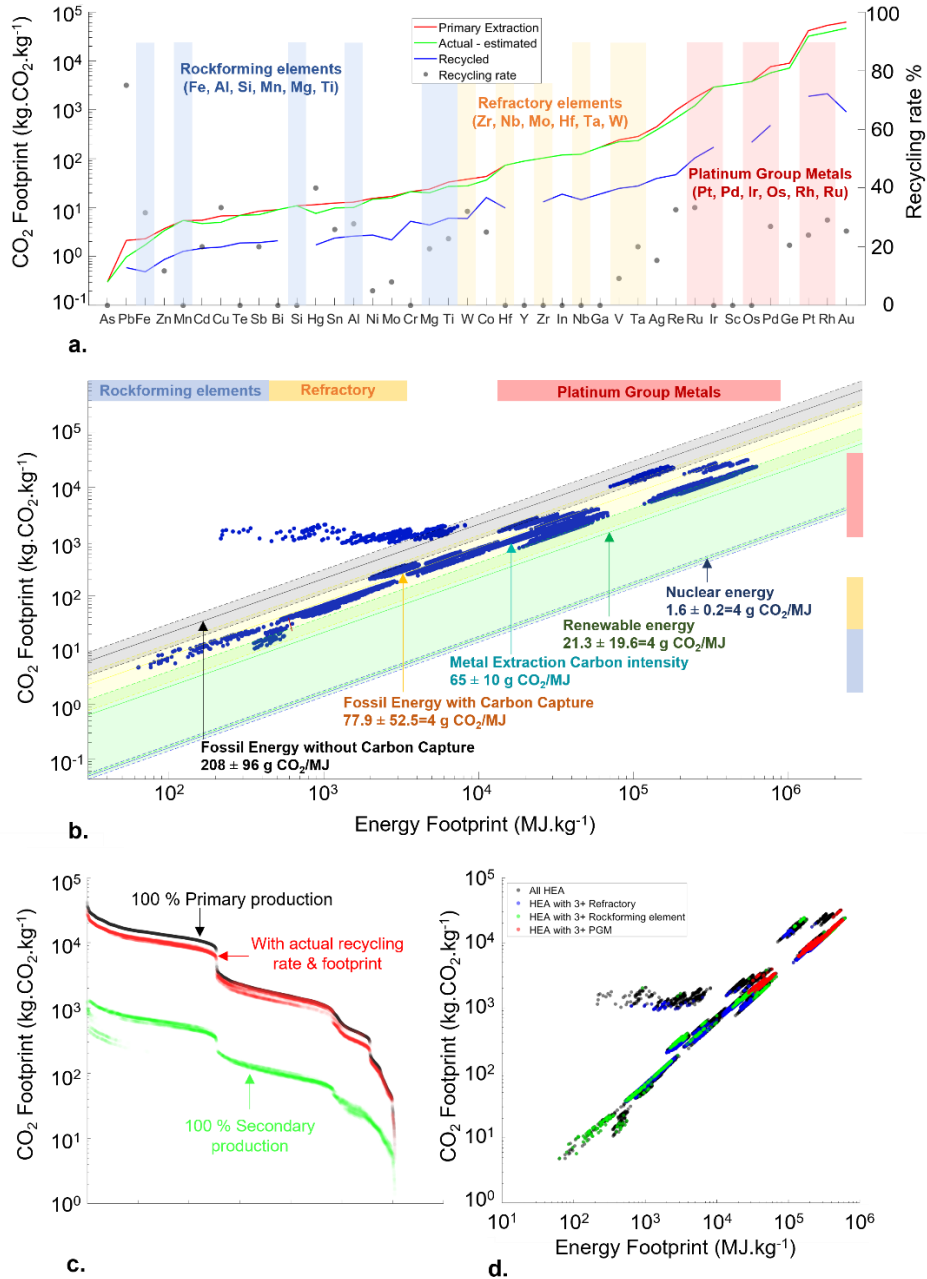

**Figure 1 | Embodied energy, CO₂ footprint, and recycling potential of metals. a**, $CO_2$ footprint versus recycling rate for major elements. Rock-forming elements (blue) show relatively low footprints (<10 kg $CO_2$/kg), refractory elements (yellow) are intermediate, while platinum group metals (PGMs, red) reach several tons $CO_2$/kg (e.g., Os) (gaps mark the non-existing data points). Lines indicate the $CO_2$ intensity from primary (red), current practice (recycling-adjusted, black), and fully recycled (green) pathways. **b**, Comparative energy and $CO_2$ footprints for element production, contextualized against energy sources. Metal extraction exhibits an average intensity of ~65 g $CO_2$/MJ (green band), but individual elements span five orders of magnitude. This range

overlaps with nuclear and renewables at the low end, and fossil fuels at the high end, showing why energy source choice is decisive. **c**, Recycling pathways illustrate emissions reductions from current practice (red line) to full secondary production (green line). **d**, Compositional clustering of HEAs: alloys rich in PGMs occupy the high-impact regime, rock-forming-element-rich alloys cluster at the low end, and refractory-rich alloys fall in between.

There is a strong correlation between energy consumption and $CO_2$ emissions in metal extraction, with an average carbon intensity of 65 ± 10 g $CO_2$/MJ (**Figure 1b,** shown by the central green band). The 30201 calculated HEAs follow this linear trend across four to five orders of magnitude in both energy use and carbon emissions. The visualization further contextualizes environmental impact of metal extraction by comparing it to various energy sources: nuclear energy (1.6 ± 0.24 g $CO_2$/MJ) and renewables (21.3 ± 19.8-4.6 g $CO_2$/MJ) exhibit significantly lower carbon intensities, while fossil fuels with carbon capture (77.6 ± 52.5-4 g $CO_2$/MJ) are comparable to current metal production practices. In contrast, conventional fossil fuels without capture (208 ± 86 g $CO_2$/MJ) result in substantially higher emissions, showing that transitioning metal extraction to low-carbon energy sources could reduce $CO_2$ emissions by factors of 3 to 40, depending on the technology adopted.

The potential impact of recycling on carbon emissions across the metal spectrum is illustrated on **Figure 1c**. Three distinct production pathways are compared: 100% primary production (black line), current practice with actual recycling rates (red line), and 100% secondary production (green line), to show the substantial emissions reduction potential through increased recycling. **Figure 1d** extends the analysis by categorizing HEAs according to their elemental composition, revealing distinct sustainability clusters. The plot compares HEAs containing three or more refractory elements (blue), rock-forming elements (green), or PGM elements (red) against the full dataset. These compositional groupings form separated energy-carbon footprint profiles, with PGM-rich alloys clustering in the high end of both metrics, while rock-forming-element-rich alloys occupy the lower end. This classification provides crucial insights for materials designers, demonstrating how strategic elemental selection can dramatically influence the alloy sustainability profile, with differences spanning several orders of magnitude in both energy consumption and carbon emissions.

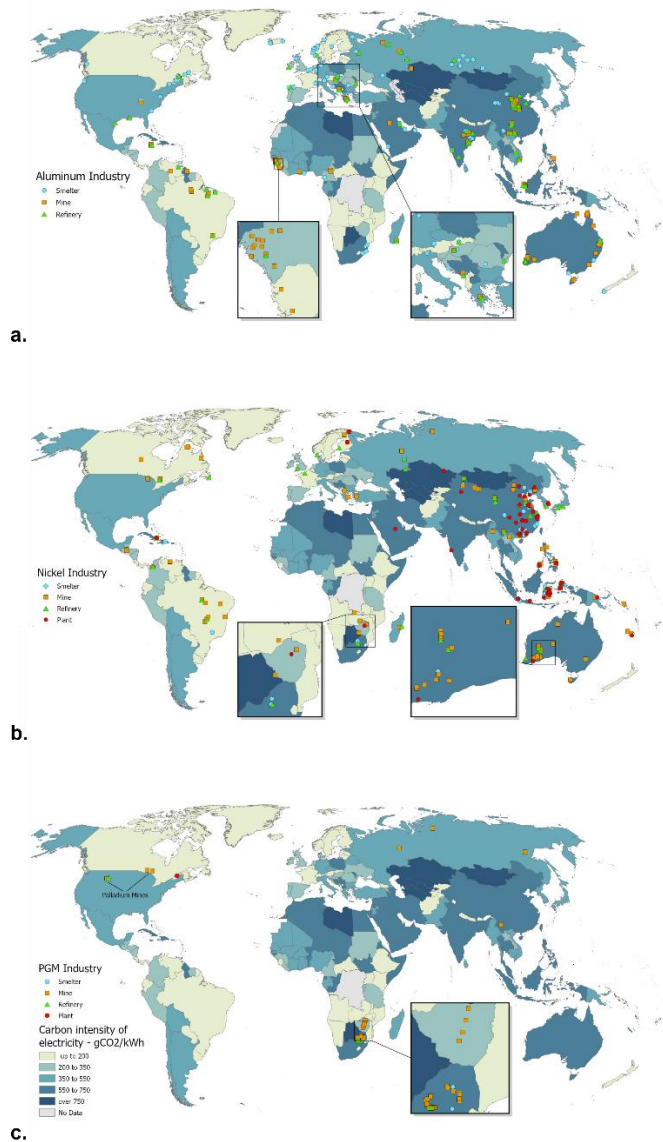

**Figure 2 | Geographical variation of CO$_2$ intensity in metal production.** Global distribution of a) aluminium, b) nickel, and c) platinum group metals (PGM) industries (smelters, mines, refineries, and plants), overlaid with regional electricity carbon intensity (gCO$_2$/kWh). Insets highlight high-density production zones. The same metal can exhibit dramatically different CO$_2$ footprints depending on where it is produced: aluminium smelting in hydro-dominated grids (e.g., Norway, Canada) yields far lower emissions than in coal-dependent grids (e.g., Poland, China, Australia). The location dependence explains why identical energy consumption can result in order-of-magnitude differences in CO$_2$ footprint, a factor that must be considered in sustainability-focused alloy design.

**Figure 2** adds a geographical perspective to the environmental impact of metal production by mapping the global distribution of the aluminum (2a), nickel (2b), and PGMs (2c) industries against the regional carbon intensity of electricity generation. These maps reveal how the same metal, when produced in different locations, can have dramatically different carbon footprints solely due to variation in local energy mixes. For instance, aluminum-smelting operations in regions powered predominantly by hydroelectricity emit significantly less $CO_2$ than those in coal-dependent regions. Therefore, the regional energy infrastructure has a pivotal role in determining the sustainability of metal production. The location-specific energy policies directly influence metal sustainability profiles and its overall environmental impact.

**Environmental, Social and Governmental (ESG) impact of mining**

Beyond carbon and energy metrics, mining operations present a range of multifaceted environmental and societal impacts. Jowitt *et al.*[25] suggest that ecological, social, and governance (ESG) factors are likely to become the dominant constraint on the metals and minerals supply in the forthcoming decades, potentially exceeding the impact of physical resource depletion.

To evaluate these risks, we draw from comprehensive assessment developed by Lebre *et al.*[2], which systematically analysed more than 6888 mining projects worldwide. This dataset covers a wide spectrum of risk factors, including waste management practices, water resource utilization, biodiversity loss, community impact, social vulnerability, land access, and governance frameworks. As a result, a composite risk score (ESG) has been derived for each country and, consequently, for each metal (see Section III and IV of the Supplementary Information). Notably, certain metals, including Gallium, Cobalt, Tantalum, and selected Platinum Group Metals (Ir, Ru, Os), evidence notably elevated ESG risk scores.

**Figure 3** visualises the distribution of HEA compositions across two axes: ESG risk and supply concentration (Herfindahl-Hirschman Index - HHI). The plot identifies three market concentration zones: highly concentrated markets (HHI > 0.25, red) containing 29403 alloys; moderately concentrated markets (0.15 ≤ HHI ≤ 0.25, yellow) with 788 alloys, and unconcentrated markets (HHI < 0.15, green) with just 10 alloys. The vast majority of calculated HEAs fall within the highly concentrated market category (red zone), indicating significant supply chain vulnerability for most potential compositions. Only 10 compositions qualify as truly diversified.

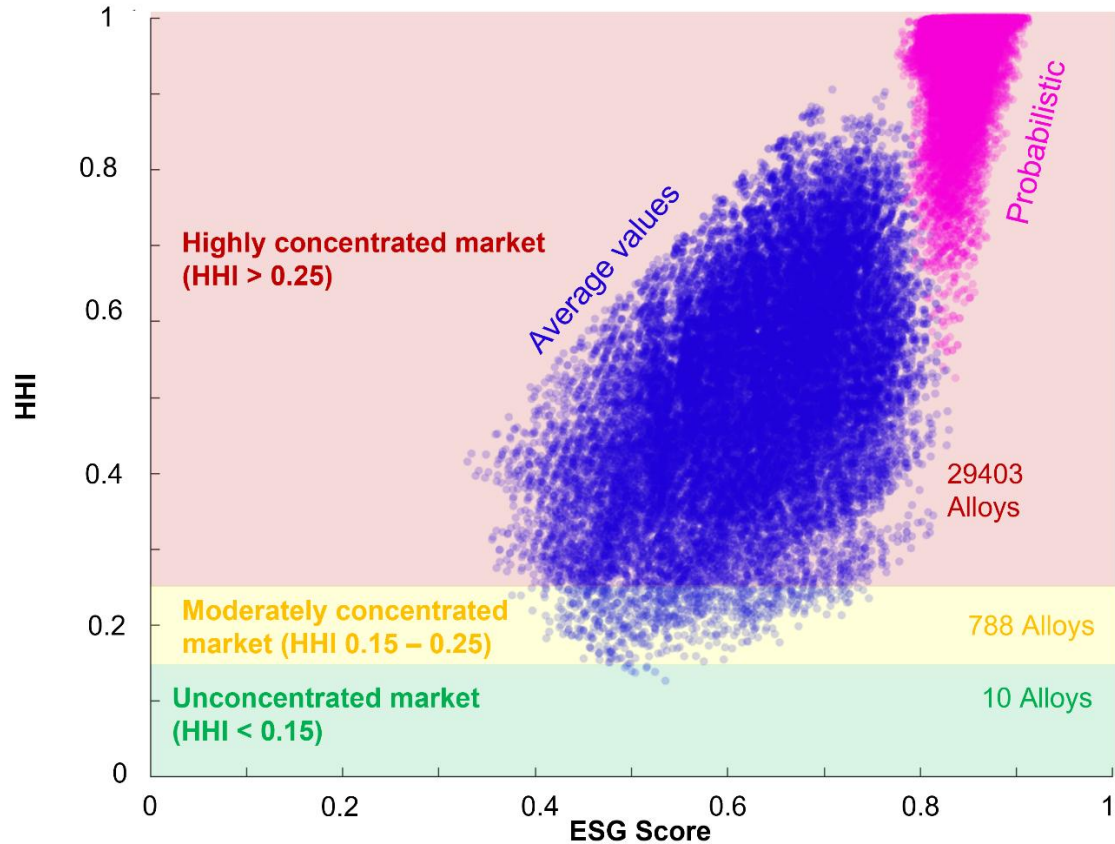

**Figure 3 | Supply chain vulnerability landscape: ESG risk versus market concentration.** Scatterplot of 30,201 equimolar HEAs across two orthogonal axes: Environmental, Social, and Governance (ESG) risk (horizontal) and supply concentration measured by the Herfindahl–Hirschman Index (HHI) (vertical) (see Methods). The plot identifies three market categories: highly concentrated (HHI > 0.25, red zone), moderately concentrated (0.15–0.25, yellow), and unconcentrated (HHI < 0.15, green). The vast majority of HEAs fall into the high-HHI, high-ESG quadrant, indicating strong vulnerability to supply disruptions. The combined supply risk metric show that most of the PGM and Ga cluster in the high-HHI-high-ESG risk zone, indicating substantial vulnerability to supply disruptions and rendering them unsuitable for large-scale industrial deployment. In contrast, alloys composed of elements with diversified supply sources and low ESG risk, such as AgAuCdCuZn, are the most stable and secure compositions in the dataset in terms of overall supply risk.

It is important to note that ESG and HHI risks behave as compounding probabilities and not as simple averages and can be interpreted as probabilities of supply disruption, consistent with a the probabilistic framework[26]. Because each alloy combines multiple elements, the overall risk is governed by the most vulnerable component. This "weakest link" principle implies that an alloy containing even a single high-risk element inherits much of that risk, regardless of the stability of its other constituents. Consequently, the alloy design strategies ~~need~~ require a fundamental shift: rather than averaging ~~out~~ risk, researchers must identify and avoid bottleneck elements altogether, considering of each constituent rather than relying on averaged risk profiles.

To consolidate supply risk assessment, we define a combined "supply-risk" metric that integrates both Environmental, Social and Governance (ESG) factors and supply assessed by the Herfindahl-Hirschman Index (HHI) (See Supplementary file).

**Companionability and by-product dependence**

In addition to the impacts associated with mining, a peculiar aspect of metal supply is "companionability", which refers to the proportion of a metal's production that co-occurs with the production output of another metal. Companionability can manifest either as a co-product, where the values of the involved metals are relatively similar, or as a by-product, where there is a significant disparity in value between the companion metal and the host metal. This concept has been quantified in earlier[20]. Our contribution is to bring companionability into the HEA design context, where it remains under-applied.

Currently, approximately 61% of naturally occurring chemical elements have companionability levels exceeding 50% (**Fig. 4**). This phenomenon is particularly notable among several PGMs (Ru, Rh, Pd, Ir, Os), primarily secured as by-products in platinum extraction. A similar pattern is observed with metals like Indium and Gallium, which are exclusively obtained as by-products of zinc and aluminium extraction.

The prevalence of companionability in metal supply raises two notable concerns. Firstly, it intertwines the demand and supply dynamics of the by-product metal with the production of the host metal. As a result, the availability of the by-product metal becomes closely linked to the demand for the host metal rather than driven by the dynamics of the by-product itself. Secondly, with the anticipated increase in metal recycling rates in the future circular economy, there is a foreseeable decrease in the proportion of metals derived from primary production by 2035-2040. As recycling is absolutely necessary to ensure supply sustainability and reducing $CO_2$ emissions, an unintended consequence may be a decline in the availability of by-product metals, posing a critical challenge for meeting future demand, unless proactive substitution strategies are developed and implemented.

Figure 4b quantifies the dependency networks between HEAs and sixteen critical host metals, revealing three distinct supply vulnerability pathways: direct dependence (blue bars, alloys containing the host metal itself), complete indirect dependence (red bars, alloys containing companion metals obtained exclusively as by-products of the host), and partial indirect dependence (green bars, alloys containing companion metals partially derived from the host metal). This detailed stratification analysis illuminates complex supply vulnerabilities that traditional criticality assessments often overlook.

The data reveals striking asymmetries in the materials ecosystem – platinum emerges as the most critical node in the HEA network, forming the highest number of alloy compositions (over 24900 combined), with approximately 6550 alloys containing platinum directly and another 18350 alloys dependent through its companion elements. Other key host metals, such as copper (18337 total dependent alloys), nickel (14180), and iron (12082), play similarly foundational roles in the supply chain, and their availability governs not only their direct use but also the feasibility of a vast number of seemingly unrelated HEA systems. This multi-layered dependency reveals systematic supply risks where disruptions propagate far beyond the affected host metal. For example, a disruption in platinum production would affect 21.7% of all calculated HEAs directly, while simultaneously restricting companion metal availability for thousands more alloys that contain no platinum themselves. Such cascading vulnerabilities create hidden systematic risks that must be considered when designing sustainable substitution strategies.

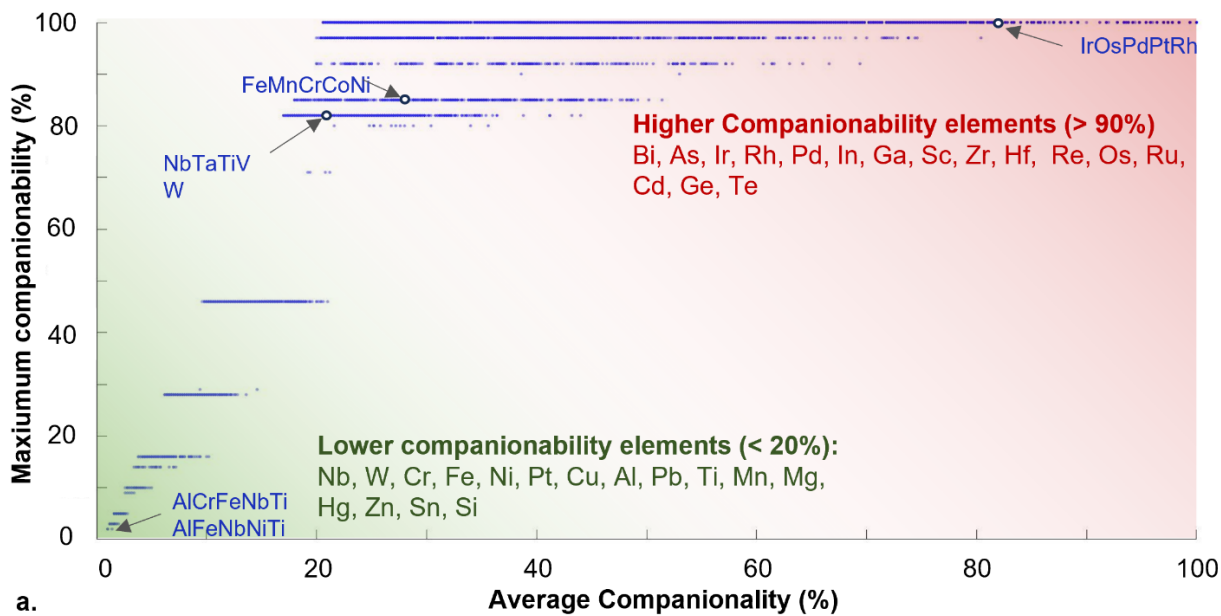

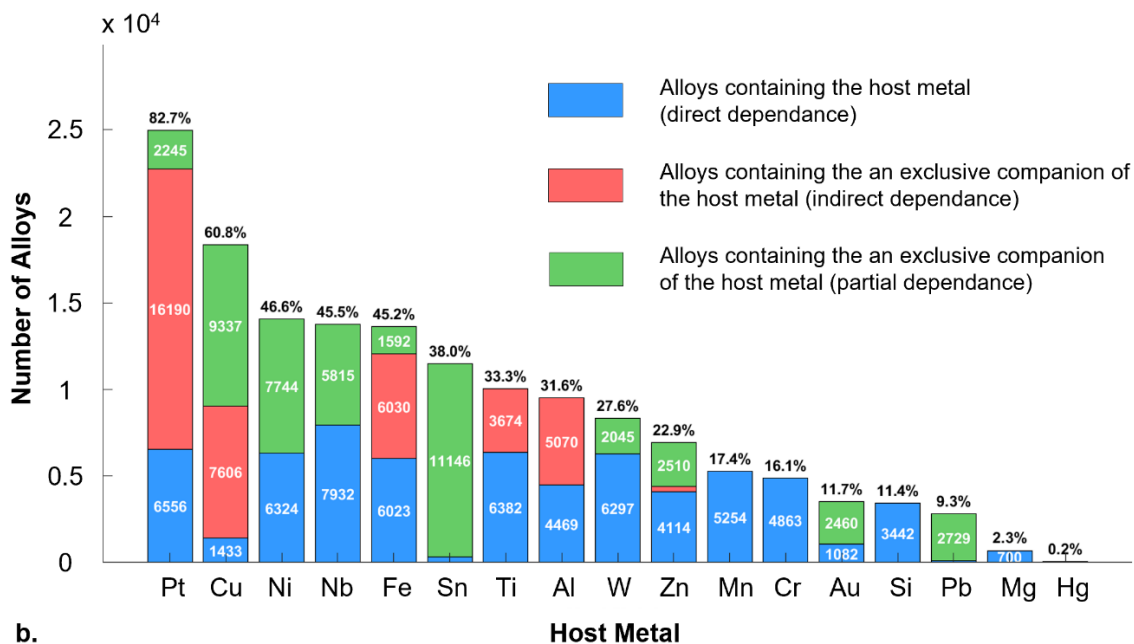

**Figure 4| Companionability and by-product dependence of HEAs a**, Distribution of average versus maximum companionability scores for the 30,201 equimolar HEAs. Average companionability captures diffuse dependence across multiple host metals, while maximum companionability highlights the single most constraining host (the "weakest link") that caps scalability. The divergence between the two metrics reveals alloys where one host element dominates feasibility despite a relatively benign average. **b**, Dependency networks between HEAs and 16 critical host metals, disaggregated into direct dependence (blue bars, alloys containing the host metal itself), complete indirect dependence (red bars, alloys containing by-product elements

exclusively tied to the host), and partial indirect dependence (green bars, alloys partially dependent on the host). Platinum emerges as the most critical node, with ~24,900 dependent alloys (6,550 directly containing Pt and ~18,350 dependent through Pt-derived by-products). Similar cascading dependencies are observed for Cu, Ni, and Fe, showing their systemic role in HEA feasibility.

In the context of HEAs, our analysis reveals that a substantial majority, 91.8% of the compositions (27719 HEAs), contain at least one element with a companionability level exceeding 95%. In stark contrast, only 1.8% of the dataset (534 HEAs), consists exclusively of elements with companionability levels below 30%.

**Comparing forecasted demand with production and reserves to avoid dead-ends**

The issue of companionability naturally extends to the broader question of the overall metal availability, whether from primary extraction or as by-products. **Figure 5** presents a comprehensive analysis of the production constraints on HEA viability through three complementary visualizations that reveal critical insights beyond simplistic abundance correlations.

**Figure 5a** maps ~~the~~ all predicted HEA compositions within a production-possibility landscape with distinct viability zones, defined by maximum yearly production capacity versus long-term sustainable production based on known reserves. Maximum annual production potential for each HEA is calculated using the formula:

$$P_{max}^{HEA} = min\left\{\frac{P_i}{x_i^{HEA}}\right\}$$

where $P_i$ is the global annual production of element $i$, and $x_i^{HEA}$ is the molar concentration of element $i$ in the alloy. This method identifies the most limiting element in each composition, setting the upper bound on feasible production volume. The analysis reveals a stark attrition curve: while 16078 compositions are viable for small-scale applications (ranging from 0.1 to 10k ton/year), only 18 compositions can support production volumes exceeding 1 million tons annually. Looking further ahead, the analysis incorporates year 2050 as a reference point aligned with major sustainability and climate policy horizons (horizontal arrow in Figure 5a). Under current reserve constraints, a broader but still narrow set of 484 compositions (1.6% of calculated HEAs) could support production volumes over 25 million tons. These calculations do not account for potential increase in reserves from future exploration or lower cut-off grades, nor do they reflect the anticipated intensification of metal recycling, ~~that~~ which ~~can both~~ could expand production feasibility over time.

**Figure 5b** presents the non-linear relationship between production thresholds and the number of viable HEA compositions. The curves, plotted for both current annual production and maximum production capacity by 2050, show "supply cliffs", i.e. critical thresholds where small increases in required production volume cause a disproportionate drop in the number of feasible alloys. This phenomenon underlines a key vulnerability, where a composition viable in a current context may become unsustainable with only a marginal increase in the production demand, often due to the bottleneck imposed by a single limiting element.

**Figure 5c** further identifies the specific elements responsible for such production constraints. A radar plot visualization compares element frequency as a limiting factor within the whole data set and within the subset of HEA containing the given element. Osmium emerges as the most restrictive, constraining 7423 compositions to extremely limited production of just ~1 ton annually. Other PGMs such as ruthenium, iridium, and rhodium similarly constrain thousands of compositions, creating inherent limitation on their practical application scale regardless of their performance characteristics. This analysis allows for pre-emptive exclusion of elements that lead to "dead-end" alloy designs, i.e. development of materials that, regardless of their functional properties, cannot be deployed at industrial scales due to hard production limits. Identifying and avoiding these critical constraints early, researchers can prioritize HEAs with realistic scale-up potential, when material discovery is guided with industrial applicability in mind.

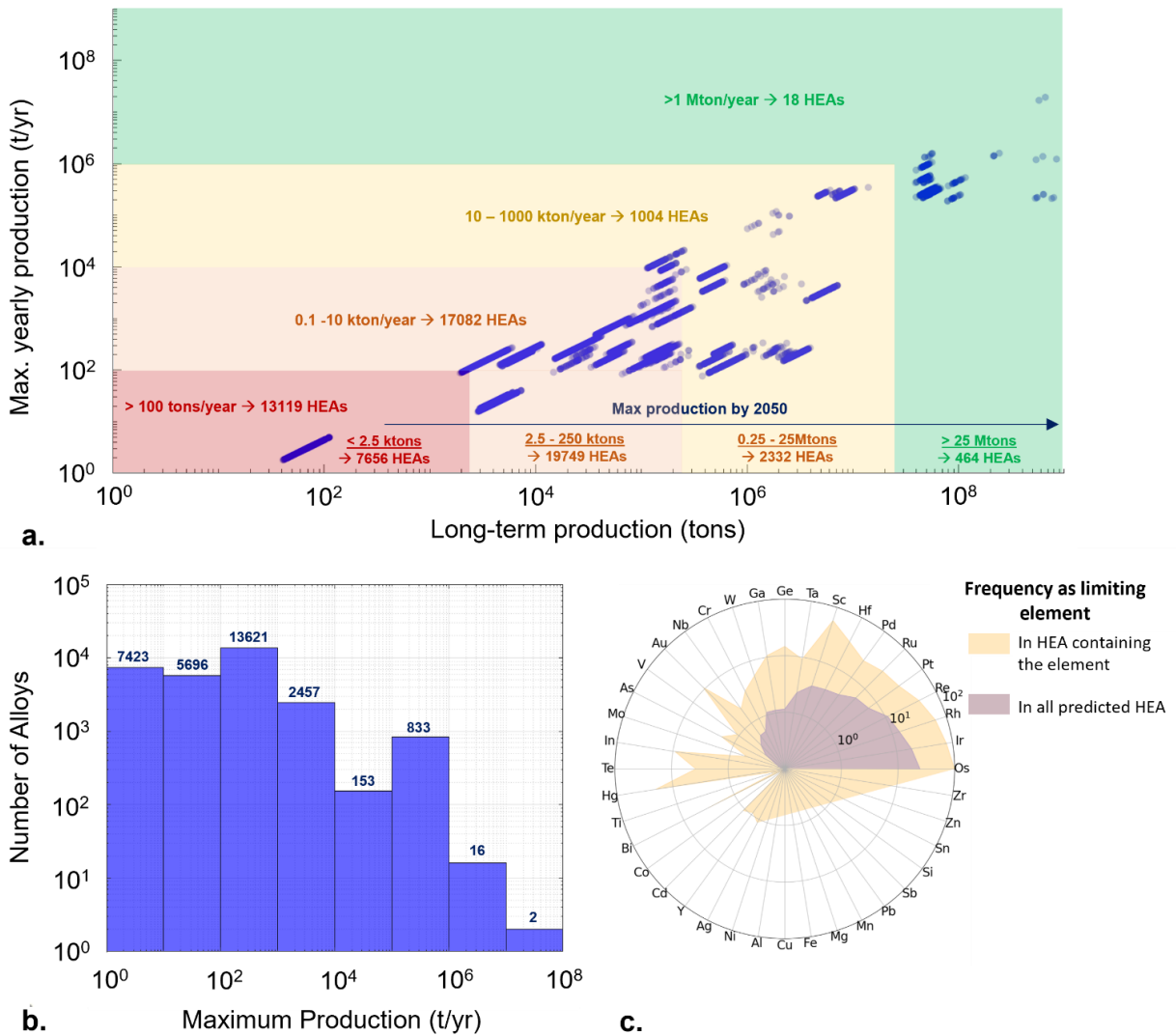

**Figure 5 | Production and Reserve Constraints on HEA Scalability. a**, Production-possibility landscape mapping maximum annual production potential (vertical axis) versus long-term sustainable production potential based on known reserves (horizontal axis). Each point represents an equimolar HEA composition; distinct feasibility zones emerge, with options narrowing dramatically as target scale increases. For example, while >16,000 HEAs could support small-scale applications (0.1–10 kton/year), only 18 compositions remain viable for >1 Mt/year production. The 2050 horizon is included as a policy-relevant benchmark. **b**, Non-linear "supply cliffs" showing the sharp drop in viable alloy numbers at certain production thresholds. Even small increases in required volume eliminate disproportionately large subsets of HEAs, often due to a single critical element. **c**, Element-specific bottlenecks identified by comparing frequency of limiting elements across the dataset. Osmium emerges as the most restrictive (≈1 t/year production), constraining >7,400 HEAs. Ruthenium, iridium, and rhodium impose similar ceilings. While demand growth can drive supply expansion where reserves exist, by-product constraints (e.g. osmium at ~1 t/year) cannot be overcome by demand alone.

Page **14** of **26**

**Gather the risks in a Sustainability Risk Index**

The advancement of sustainable materials design requires robust and transparent decision-making support systems. While Life Cycle Assessment (LCA) can provide high accuracy for mature technologies by integrating synthesis pathways, geographical factors, energy carbon footprints, and transportation considerations, its complexity limits scalability to vast theoretical materials databases. To address this gap between computational alloy discovery and experimental realization, we propose a streamlined multi-criteria decision-making (MCDM) framework for the preliminary screening of candidate materials.

Our methodology comprises three key stages. First, parameter clustering was done by Pearson correlation analysis (**Figure 6a** and Supplementary Information), revealing four distinct parameter groups with strong internal correlations: (i) environmental impact metrics ($CO_2$ footprint and embedded energy), (ii) companionability indicators, (iii) material availability metrics (production capacity and reserves), and (iv) supply chain factors (supply risk, Herfindahl-Hirschman Index (HHI), and Environmental, Social, and Governance (ESG) scores). Second, we implemented two complementary criteria weighting strategies. The entropy method (See Supplementary Information) objectively assigns weights based on data dispersion across alloy candidates, and this approach was implemented for both averaged and probabilistic companionability and supply risk assessments. Further, we developed scenario-based user-defined weighting schemes representing: (1) Equi-proportional criteria importance, (2) long-term sustainability prioritization (predominance on $CO_2$ and Reserves), and (3) short-term implementation feasibility (predominance of $CO_2$, Production and Supply risk). The five resultant weighting distributions are represented in **Figure 6c**. Finally, we established a normalized scoring system wherein alloy performance for each criterion was transformed to a decile-based scale (0.1-1.0), with an alloy placing in the 5th decile receiving a score of 0.5 for that criterion. These normalized scores were then multiplied by their respective criteria weights for each scenario; **Figure 6b** presents the resulting sustainability risk distributions

**1556 HEAs have high sustainability score regardless the criteria allocation**

By setting a maximum sustainability risk score at 0.25, we observe that between 6.5% and 9.9% of the evaluated HEAs meet this sustainability criterion, depending on which weighting methodology is applied. The histograms in **Figure 6b** clearly illustrate this distribution, with the green-shaded regions highlighting alloys below the 0.25 threshold.

Notably, a subset of 1556 HEAs maintains a sustainability risk score below 0.25 regardless of criteria weighting, representing approximately 5% of the entire data set. These "resiliently sustainable" alloys are particularly valuable, as their favourable sustainability profile remains stable regardless of shifts in assessment priorities. As shown in **Figure 6c**, the five different criteria weighting schemes place significantly different importance on key criteria such as supply risk, companionability, $CO_2$ emissions, production capacity, and reserves. Despite these substantial variations in how sustainability is evaluated, either from probabilistic risk assessment to long-term focused approaches, these 1556 alloys consistently rank among the most sustainable candidates. This resilience to methodological variations suggests that these HEAs possess inherently favourable characteristics across multiple sustainability dimensions simultaneously, rather than excelling in just one or two categories. As such, they are prime candidates for further experimental validation and potential industrial application, as they offer sustainability advantages that are resilient to evolving material selection frameworks and policy-driven priorities.

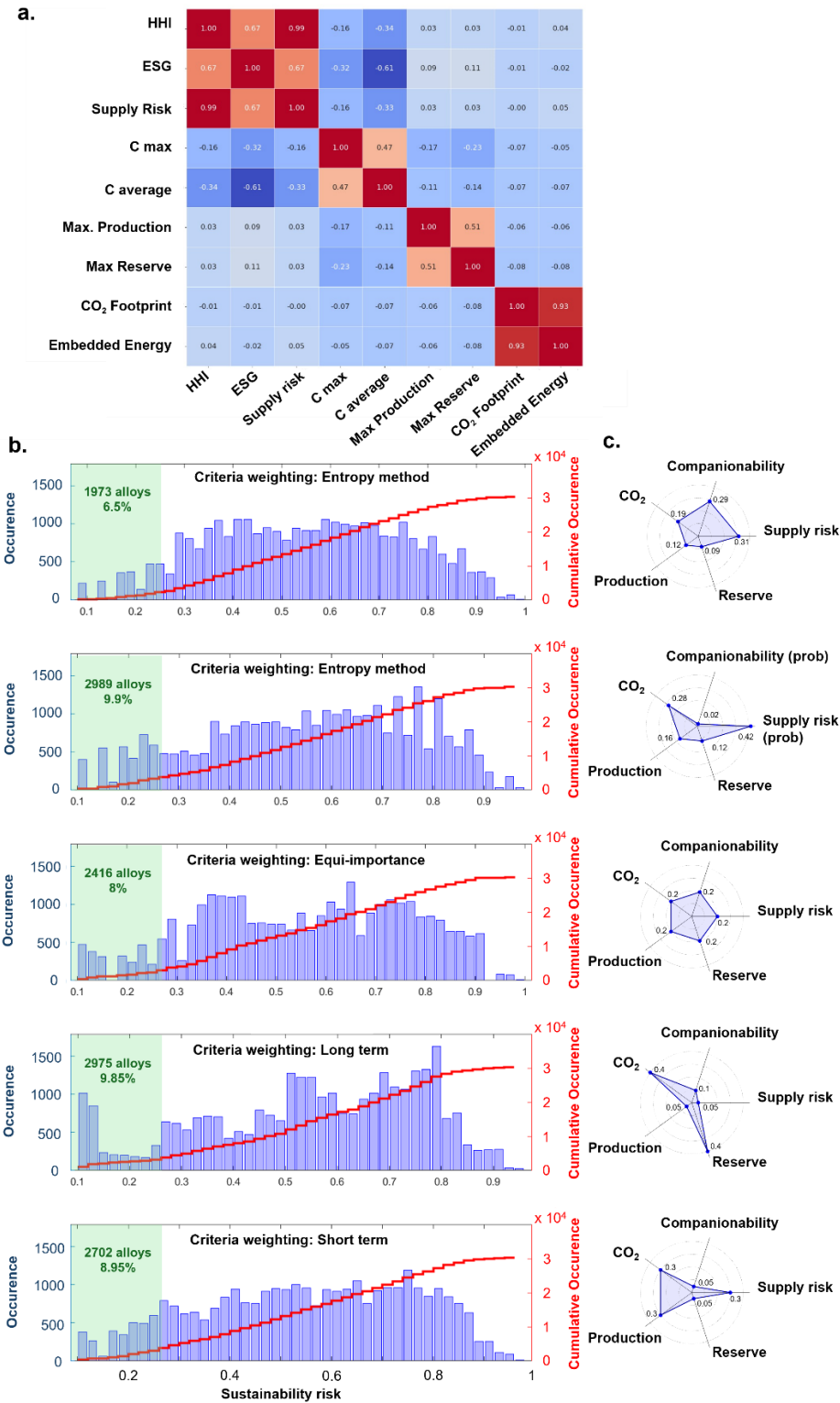

**Figure 6 | Consolidating sustainability risks into a composite index.** Pearson correlation matrix showing clustering of sustainability indicators into four groups: (i) environmental impact ($CO_2$ footprint, embedded energy), (ii) companionability metrics, (iii) availability metrics (production

capacity, reserves), and (iv) supply chain risks (supply risk, Herfindahl-Hirschman Index (HHI), and Environmental, Social, and Governance (ESG) scores). **b**, Distribution of sustainability risk scores for the 30,201 HEAs under five complementary weighting schemes. The histograms show that 6.5–9.9% of alloys fall below the 0.25 sustainability risk threshold, depending on weighting. A resilient subset of 1,556 alloys consistently remains below this threshold across all schemes, representing ≈5% of the dataset. **c**, Five weighting families tested: entropy-based (objective weighting from data dispersion), equiproportional (equal weights), long-term prioritization ($CO_2$ and reserves emphasized), short-term feasibility (production and supply risk emphasized), and probabilistic/weakest-link approaches. Despite strong differences in weight allocations, the resilient subset is stable across all methods.

**HEAs are not a suitable substitute for the Stainless steel in the mass market**

The importance of weighing the equilibrium between the performance of materials and their practical feasibility has garnered increased attention[27]. Consequently, researchers are encouraged to broaden their evaluation criteria beyond traditional performance metrics, and placing emphasis also on the scalability and real-world applications of new materials.

Although the full assessment of the mechanical and functional performance of HEAs is beyond the scope of this study, it is essential to evaluate their sustainability and scalability in alignment with anticipated application requirements. For structural applications, particularly as potential substitutes for stainless steel, the required production volume reaches tens to hundreds of millions of tons annually. Under such conditions, raw material availability becomes a major limiting factor for HEAs practical implementation. Conversely, applications like catalysts for hydrogen production require only hundreds of tons per year (see **Figure 7**). To assess practical feasibility, we conducted a cross-analysis of each alloy that combines the Sustainability Risk Score with maximum production capacity, derived from the production solicitation data presented in **Figure 5**.

A multidimensional analysis of the feasibility of HEAs across several different market segments through five complementary visualizations are presented in **Figure 7**. More specifically, **Figure 7a** presents a heatmap showing the number of eligible alloys at each intersection of targeted production volume and sustainability risk threshold, revealing distinct "opportunity zones" where viable substitutions exist. **Figure 7b** maps specific HEA compositions against these constraints, highlighting exemplar alloys like 304 stainless steel, Titanium and Platinum at their respective production scales. The distribution of blue points shows how rapidly options diminish as production requirements increase.

To provide deeper insights into element-specific constraints the alloys containing critical bottleneck elements like osmium (**Figure 7c**), iridium (**Figure 7d**), and hafnium (**Figure 7e**) are highlighted. These visualizations reveal how certain elements systematically constrain production potential regardless of sustainability score: osmium-containing alloys (red points in **Figure 7c**) are limited to extremely low production volumes, while iridium and hafnium similarly impose distinct production ceilings due to their limited global availability.

HEAs have demonstrated promising mechanical properties, both at high-temperature and cryogenic environments. As a practical case study, we assess the feasibility of HEAs as potential replacements for 304 stainless steel alloy, a widely used material in cryogenic applications. Our search has yielded a limited pool of substitution options since only a few HEA systems were identified as potential alternatives for only a fraction of the stainless steel production (**Fig. 7b**). Indeed, 304 stainless steel alone accounts for approx. 60 million tons annually, but is yet just a subset of the global steel production exceeding 2 billion tons per year. Despite the broad compositional space assessed, no HEA composition was found capable of supporting production at this scale, regardless of its sustainability risk score, showing a fundamental scalability limitations for mass-market structural applications, despite their potential superior material properties.

While this study provides a sustainability-screening framework at the alloy design stage, we acknowledge that manufacturability and processing routes are decisive for eventual industrial deployment. Processing energy demand, recycling efficiency, and location-specific electricity mixes can all significantly influence sustainability outcomes. These aspects, however, are highly process- and country-dependent and thus fall outside the scope of this computational screening. We emphasize that our results are intended as a pre-selection compass for experimental validation, after which detailed process-specific assessments should follow. This design philosophy parallels other materials domains such as batteries and magnets, where slightly less high-performing but more supply-secure materials are increasingly preferred over high-performing but supply-constrained options.

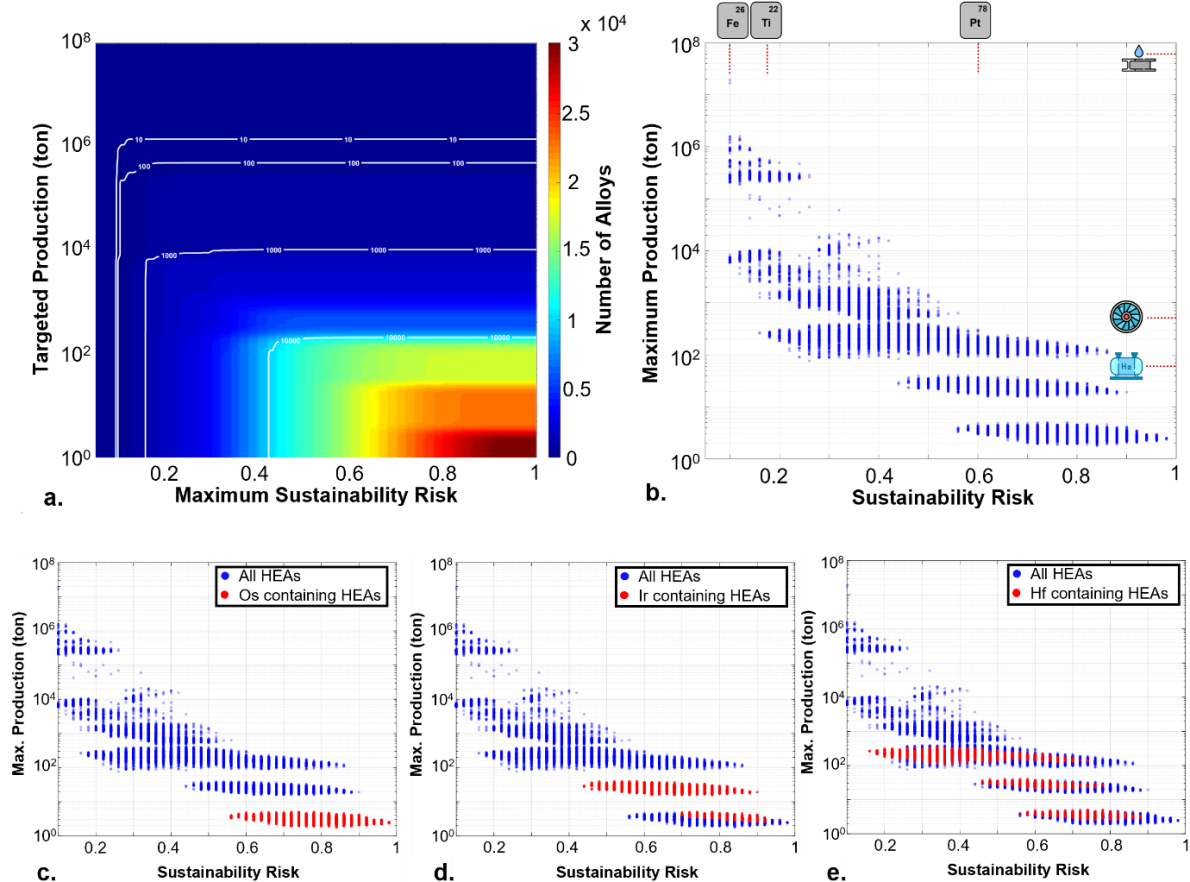

**Figure 7 | Feasibility of HEAs across different application scales. a**, Heatmap of possible HEA compositions at each intersection of sustainability risk score and targeted production volume. Distinct "opportunity zones" emerge, where viable substitutions exist (low-risk alloys at modest production scales). **b**, Comparison of HEAs against widely used reference alloys. Examples include 304 stainless steel (~60 Mt/year), Grade 5 Ti alloy (~150 kt/year), and platinum (~200 t/year). HEAs with low sustainability risk and sufficient production potential cluster as promising substitutes for catalytic and niche structural applications, whereas no HEA composition can feasibly replace mass-market steels at multi-megaton scales. **c–e**, Bottleneck constraints highlighted by isolating alloys containing specific limiting elements: osmium (c), iridium (d), and hafnium (e). Alloys containing these elements are shown in red, demonstrating how single elements impose hard ceilings on feasible production regardless of sustainability risk. This figure shows the natural niches where HEAs are most promising: thousands of candidates are viable at catalytic and niche structural scales, but fundamental resource constraints prevent their substitution of mass-market steels.

### Thousands of candidates to substitute Platinum for catalysis

The use of High-Entropy Alloys (HEAs) in catalysis, especially in hydrogen production, is a relatively recent yet promising development. Given the expected increase in demand for hydrogen production via water electrolysis, a projected shortfall of approximately 100 tons of platinum is expected by 2030-2040, whereas the current production rate is around 200 tons annually. In this context, HEAs offer a compelling opportunity for replacing platinum (**Fig. 7b**). Considering critical constraints outlined in Fig. 7a and 7b, we identified a subset of 15194 HEAs capable of achieving production volumes up to 100 tons, making them sustainable candidates to substitute platinum in catalytic applications. This subset represents 50.3% of the total examined dataset. It is important to note that while these HEAs may not outperform Platinum Group Metals (PGM) in catalytic activity, they meet the critical threshold for industrial feasibility, which is a rare and valuable property given the scarcity of viable PGM alternatives, revealing a broad compositional landscape with significant potential for development of sustainable, scalable catalysts.

**A new era of HEA exploration with a new compass**

This study presents a sustainability-screening framework for 30,201 computationally predicted HEAs, integrating $CO_2$ footprint, energy use, ESG impacts, companionability, production capacity, and reserve availability. The developed framework provides a strategic compass to guide ~~ing~~ future HEA experimental exploration and research.

First, HEAs show remarkable application-specific sustainability profiles. While thousands of compositions (15194 or 50.3%) could potentially serve as sustainable platinum substitutes for catalytic applications at production scales around 100 tons, virtually no HEA compositions can sustainably replace structural steels at their required million-ton production scales. This production-scale sustainability relationship creates natural application niches where HEAs can make their greatest impact.

Second, the "weakest link" principle in multi-element systems means that including even one critical element with high sustainability risks compromises the entire alloy, emphasizing the strategic importance of element selection over composition optimization when designing sustainable materials. Our analysis demonstrates that avoiding platinum group metals and elements with high companionability leads to HEAs with dramatically improved sustainability profiles.

Our sustainability-based mapping of the HEA composition space empowers materials scientists to align performance goals with resource realities. It provides a valuable rapid pre-screening decision-making framework for targeting compositions that are not only functionally advanced but also scalable and environmentally responsible. Our contribution goes beyond dataset scale: (i) a feasibility map that reveals sharp supply cliffs, (ii) quantified companionability networks exposing hidden bottlenecks, and (iii) a resilient shortlist (~5%) robust across weighting schemes. Together, these provide actionable rules for materials scientists to pre-screen HEAs for sustainability prior to experimental validation.

**References**

1. Vidal, O., Goffé, B. & Arndt, N. Metals for a low-carbon society. *Nature Geoscience* **6**, 894–896 (2013).
2. Lèbre, É. *et al.* The social and environmental complexities of extracting energy transition metals. *Nat Commun* **11**, 4823 (2020).
3. Watari, T. *et al.* Total material requirement for the global energy transition to 2050: A focus on transport and electricity. *Resources, Conservation and Recycling* **148**, 91–103 (2019).
4. Zeng, A. *et al.* Battery technology and recycling alone will not save the electric mobility transition from future cobalt shortages. *Nat Commun* **13**, 1341 (2022).
5. Ali, S. H. *et al.* Mineral supply for sustainable development requires resource governance. *Nature* **543**, 367–372 (2017).
6. Castillo, E., del Real, I. & Roa, C. Critical minerals versus major minerals: a comparative study of exploration budgets. *Miner Econ* (2023) doi:10.1007/s13563-023-00388-w.
7. Metals for Clean Energy. https://eurometaux.eu/metals-clean-energy/?5.
8. Watari, T., Nansai, K. & Nakajima, K. Major metals demand, supply, and environmental impacts to 2100: A critical review. *Resources, Conservation and Recycling* **164**, 105107 (2021).
9. El-Atwani, O. *et al.* Outstanding radiation resistance of tungsten-based high-entropy alloys. *Science Advances* **5**, eaav2002 (2019).
10. Gludovatz, B. *et al.* A fracture-resistant high-entropy alloy for cryogenic applications. *Science* **345**, 1153–1158 (2014).
11. Lei, Z. *et al.* Enhanced strength and ductility in a high-entropy alloy via ordered oxygen complexes. *Nature* **563**, 546–550 (2018).
12. Yao, Y. *et al.* Carbothermal shock synthesis of high-entropy-alloy nanoparticles. *Science* **359**, 1489–1494 (2018).
13. Kumar Katiyar, N., Biswas, K., Yeh, J.-W., Sharma, S. & Sekhar Tiwary, C. A perspective on the catalysis using the high entropy alloys. *Nano Energy* **88**, 106261 (2021).
14. Cantor, B. Multicomponent high-entropy Cantor alloys. *Progress in Materials Science* **120**, 100754 (2021).
15. Liu, X., Zhang, J. & Pei, Z. Machine learning for high-entropy alloys: Progress, challenges and opportunities. *Progress in Materials Science* **131**, 101018 (2023).
16. Graedel, T. E. *et al.* Methodology of Metal Criticality Determination. *Environ. Sci. Technol.* **46**, 1063–1070 (2012).
17. Graedel, T. E., Harper, E. M., Nassar, N. T., Nuss, P. & Reck, B. K. Criticality of metals and metalloids. *Proceedings of the National Academy of Sciences* **112**, 4257–4262 (2015).

18. Ryter, J., Fu, X., Bhuwalka, K., Roth, R. & Olivetti, E. Assessing recycling, displacement, and environmental impacts using an economics-informed material system model. *Journal of Industrial Ecology* **26**, 1010–1024 (2022).
19. Fu, X., Schuh, C. A. & Olivetti, E. A. Materials selection considerations for high entropy alloys. *Scripta Materialia* **138**, 145–150 (2017).
20. Fu, X., Polli, A. & Olivetti, E. High-Resolution Insight into Materials Criticality: Quantifying Risk for By-Product Metals from Primary Production. *Journal of Industrial Ecology* **23**, 452–465 (2019).
21. Fu, X., Schuh, C. A. & Olivetti, E. A. Materials selection considerations for high entropy alloys. *Scripta Materialia* **138**, 145–150 (2017).
22. Gorsse, S., Langlois, T., Yeh, A.-C. & Barnett, M. R. Sustainability indicators in high entropy alloy design: an economic, environmental, and societal database. *Sci Data* **12**, 288 (2025).
23. Chen, W. *et al.* A map of single-phase high-entropy alloys. *Nat Commun* **14**, 2856 (2023).
24. Nassar, N. T., Graedel, T. E. & Harper, E. M. By-product metals are technologically essential but have problematic supply. *Science Advances* **1**, e1400180 (2015).
25. Jowitt, S. M., Mudd, G. M. & Thompson, J. F. H. Future availability of non-renewable metal resources and the influence of environmental, social, and governance conflicts on metal production. *Commun Earth Environ* **1**, 1–8 (2020).
26. Rousseau, F., Nominé, A., Zavašnik, J. & Cvelbar, U. Is alloying a promising path to substitute critical raw materials? *Materials Today* **83**, 1–8 (2025).
27. The fine line between performance improvement and device practicality. *Nat Commun* **9**, 5268 (2018).
28. *Mineral Commodity Summaries 2022*. *Mineral commodity summaries 2022* vol. 2022 202 http://pubs.er.usgs.gov/publication/mcs2022 (2022).

**Acknowledgements:** This work has been partially supported by the European Union through the Erasmus Program ('Cooperation Partnership' Project Title: HERawS (Highlights on European Raw materials Sustainability - Project No.: 2022-1-FR01-KA220-HED-000087621). U.C. and J.Z. acknowledges the support by the Slovenian Research Agency (research core funding no. P1-0417). The Slovenian Research Agency did not design, collect, analyse, or interpret any data and was not involved in the submission decision of the authors. J.Z. acknowledges the support from Max-Planck- Gesellschaft Partner Group. DB and HH are grateful to the University of Alberta for the partial funding of DB. AEK, DB, OO, NTH, WA, AD, OK and EK are grateful to Mines-Nancy Foundation for financial support.

**Author Contributions:** A.N. and J.Z. conceived the experiments. A.E., D. B., O. O., N. T. H., W. A., A. D., O. K., E. K. C. L., A. B. conducted experiments and performed numerical simulations. All authors analysed the results. A.N., J.Z. and U.C. prepared the manuscript and figures. All authors A.E., D. B., O. O., N. T. H., W. A., A. D., O. K., E. K. C. L., A. B., H. H., T. B., E. M. A. S., V. M., O. C., A. C., M.C., U. C., J. Z., and A. N. contributed to the compilation and review of the manuscript.

**Competing interests:** The authors declare no competing interests.

**Additional Information**

**Supplementary information** is available for this paper at DOI

**Correspondence and requests for materials** should be addressed to Janez Zavašnik and Alexandre Nominé.

**Reprints and permissions information** is available at http://www.nature.com/reprints.

**METHODS**

**Carbon and Energy Footprint calculations**

Mass Carbon $F^{CO_2}$ and Energy $F^E$ footprint per unit of mass of each metal $m$ have been found in GrantaEduPack© Software. The footprints of the HEA $j$ composed of $n$ metals $m$ in a mass concentration $x$ have been calculated as follows:

$$f_{HEA}^{CO_2} = \sum_{m=1}^{n} x_m f_m^{CO_2}$$

$$f_{HEA}^{E} = \sum_{m=1}^{n} x_m f_m^{E}$$

*Supply and ESG risk*

Supply concentration is estimated using the Herfindahl-Hirschman Index $HHI$, which is expressed for a metal $m$ as follows:

$$HHI(m) = \sum_{c} s_c^2(m)$$

Producing countries $c$ and their respective production share $s_c$ are obtained from the US Geological Survey[28] (USGS). ESG risk assessment per country has been published by Lèbre *et al.*[2]. The ESG risk of each metal has been calculated as follows:

$$ESG(m) = \sum_{c} E_c s_c(m) + S_c s_c(m) + G_c s_c(m)$$

The supply risk indicator for an HEA $j$ combines the supply concentration and the ESG risk with the following formula:

$$SR_{HEA} = \sum_{m=1}^{n} \sum_{C} (E_c + S_c + G_c)\, s_c^2(m)$$

*Companionality*

The level of companionability is taken from Nassar *et al.* [24]. Two indicators representing respectively the average companionability $\overline{K}$ and the maximal companionality $K^{max}$ are calculated as follows:

$$\overline{K}_{HEA} = \sum_{m=1}^{n} x_m K_m$$

$$K_{HEA}^{max} = \max\{K_1, K_2 \ldots K_m\}$$

*Production and Reserves*

The production $P_m$ (respectively reserve $R_m$) of a metal $m$ are found in the US Geological Survey. The maximum yearly production of an HEA $P_{HEA}^{max}$ can be calculated using a "weakest link approach":

$$P_{HEA}^{max} = \min\left\{\frac{P_1}{x_1}, \frac{P_n}{x_n} \ldots \frac{P_n}{x_n}\right\}$$

A similar approach is used to have a longer term approach, using the reserve instead of the yearly production:

$$P_{HEA}^{max} = \min\left\{\frac{R_1}{x_1}, \frac{R_n}{x_n} \ldots \frac{R_n}{x_n}\right\}$$

**Weighting schemes and supply risk aggregation**

To demonstrate different approaches robustness, we implemented five complementary weighting strategies:

1. **Entropy-based weighting**; objectively derived from the dispersion of data across alloys, giving higher weight to indicators with greater variability.

2. **Equiproportional weighting**; equal weights assigned to all criteria to avoid arbitrary prioritization.
3. **Long-term sustainability weighting**; prioritizing $CO_2$ footprint and reserves to reflect large-scale decarbonisation and resource availability goals.
4. **Short-term feasibility weighting**; prioritizing production and supply risk to reflect near-term industrial deployability.
5. **Probabilistic/weakest-link weighting**; treating supply chain disruptions as compounding probabilities, such that an alloy inherits the vulnerability of its riskiest element rather than an average risk.

For each strategy, criteria were normalized to a 0.1–1.0 decile scale, and overall sustainability scores were calculated as weighted sums. The resulting distributions (Fig. 6b) show that while absolute rankings vary modestly with weighting, a resilient subset of 1,556 HEAs (≈5% of the dataset) consistently achieves a sustainability risk score <0.25 regardless of weighting approach.

This reconciliation of average, probabilistic, and weakest-link treatments shows that the prioritization of alloys is not an artifact of methodology but an inherent feature of the dataset.